# Title

Equilibrium Phase Diagrams of Isostructural and Heterostructural Two-Dimensional Alloys from First Principles

# Author List and Affiliations

John Cavin[1,*] and Rohan Mishra[2,†,‡]

[1]Department of Physics, Washington University in St. Louis, St. Louis, MO 63130, USA.

[2]Department of Mechanical Engineering and Material Science, and Institute of Materials Science and Engineering, Washington University in St. Louis, St. Louis, MO 63130, USA.

# Lead Contact Footnote

[‡]Lead contact

# Email Addressed of Corresponding Authors

Corresponding author: john.cavin@northwestern.edu

Corresponding author: rmishra@wustl.edu

# Summary

Alloying is a successful strategy for tuning the phases and properties of two-dimensional (2D) transition metal dichalcogenides (TMDCs). To accelerate the synthesis of TMDC alloys, we present a method for generating temperature-composition equilibrium phase diagrams by combining first-principles total energy calculations with thermodynamic solution models. This method is applied to three representative 2D TMDC alloys: an isostructural alloy, $MoS_{2(1-x)}Te_{2x}$, and two heterostructural alloys, $Mo_{1-x}W_xTe_2$ and $WS_{2(1-x)}Te_{2x}$. Using density-functional theory and special quasi-random structures, we show that the mixing enthalpy of these binary alloys can be reliably represented using a sub-regular solution model fitted to the total energies of relatively few compositions. The cubic sub-regular solution model captures three-body effects that are important in TMDC alloys. By comparing phase diagrams generated with this method to those calculated with previous methods, we demonstrate that this method can be used to rapidly design phase diagrams of TMDC alloys and related 2D materials.

---

[*]Corresponding author: john.cavin@northwestern.edu
[†]Corresponding author: rmishra@wustl.edu

**Introduction**

Monolayer transition metal dichalcogenides (TMDCs) form a large class of two-dimensional (2D) materials displaying a diverse array of properties, including direct band gaps (Splendiani et al., 2010), superconductivity (Saito et al., 2016), topological insulation (Qian et al., 2014), spin and valley polarization (Mak et al., 2014a; Mishra et al., 2013; Zhou et al., 2020), and charge density waves (Rossnagel, 2011; Tsen et al., 2015). TMDCs have a stoichiometry of $MX_2$, where *M* is a transition metal and *X* is a chalcogen (S, Se, and Te). The combinatorics of the available transition metal and chalcogen choices, and the variety of crystal structures TMDCs can adapt are responsible for their diverse properties. Consequently, alloying two or more transition metals or chalcogens has been a successful method for improving their performance in electronics (Hu et al., 2020; Wang et al., 2020a) and optoelectronics (Klee et al., 2015; Raffone et al., 2016; Yu et al., 2017), and as catalysts for a variety of reactions including hydrogen evolution (Gong et al., 2016), carbon dioxide reduction (Cavin et al., 2021; Hemmat et al., 2020), and oxygen evolution and reduction (Hemmat *et al.*, 2020). Not only the properties, but the alloy structure or phase is also sensitive to composition. Common polytypes of monolayer TMDCs include the *2H* phase where the transition metal is in a trigonal prismatic coordination of the chalcogen atoms, the *1T* phase where the transition metal in an octahedral coordination, and *1T′* phase with the metal atoms in a distorted octahedral coordination. For a given composition, different polytypes can have vastly different properties. For example, monolayer $MoS_2$ in the *2H* phase is a direct band gap semiconductor with a lack of inversion symmetry that imparts a valley degree of freedom (Mak et al., 2014b); its *1T* phase is metallic, which makes it attractive for electrocatalysis (Voiry et al., 2013); and its *1T′* phase is a topological insulator protected by inversion symmetry (Qian *et al.*, 2014). By alloying two TMDCs with different ground state polytypes, a process referred to as heterostructural alloying, it is possible to stabilize an otherwise metastable polytype, and obtain new functionalities (Deng et al., 2021). This method of phase engineering through heterostructural alloys has been used to stabilize the *2H* and *1T′* phases in $WSe_{2(1-x)}Te_{2x}$ (Yu *et al.*, 2017).

For isostructural and heterostructural alloys alike, identifying regions in the temperature-composition space where a single phase of the alloy is stable can accelerate their synthesis. This relationship is summarized quantitatively through equilibrium temperature-composition phase diagrams. Isostructural alloys typically have the same crystal structure as their end members, although there are exceptions (Zunger, 1997). Hence, their phase diagrams are relatively straight forward and provide a miscibility temperature above which the alloy is stable in a single phase. Such phase diagrams can be generated from thermodynamic quantities obtained using first-principles density-functional-theory (DFT) calculations (Ferreira et al., 1989; Wei et al., 1990b; Zunger, 1997). Isostructural alloys of monolayer TMDCs have received widespread attention following reports of their theoretical phase diagrams. Kang et al. reported one of the first phase diagrams of 2D chalcogen alloys of group VI transition metals having the *2H* phase (Kang et al., 2013a; Kang et al., 2013b). A recent work reported the experimental realization of several isostructural TMDC alloys involving group V and VI transition metals, with alloying at both the metal- and chalcogen-sites, using theoretically predicted phase diagrams (Hemmat *et al.*, 2020). In contrast to isostructural alloys, heterostructural TMDC alloys — where the end members have different crystal structure — are starting to garner attention only recently (Aslan et al., 2018; Li et al., 2016; Oliver et al., 2017; Oliver et al., 2020; Rhodes et al., 2017; Wang et al., 2020b). The phase diagrams of heterostructural alloys are comparatively nontrivial, having multiple phase boundaries. There are only a handful of examples of phase diagrams of heterostructural 2D TMDC alloys; many of them have been obtained using time-consuming growth experiments at different temperatures and compositions followed by characterization (Oliver *et al.*, 2017; Rhodes *et al.*, 2017). Relatively few diagrams have been predicted theoretically, such as the report on $Mo_{1-x}W_xTe_2$, a heterostructural alloy of *2H* $MoTe_2$ and *1T′* $WTe_2$, by Duerloo et al. (Duerloo and Reed, 2016).

In this Article, we present an efficient framework to generate equilibrium phase diagrams of 2D TMDC alloys that are based on a limited number of DFT calculations. We fit the mixing enthalpy of the alloys with a cubic sub-regular solution model that can capture the important interactions between the triplet clusters in TMDCs. We apply this method to generate the phase diagrams of heterostructural TMDC alloys $Mo_{1-x}W_xTe_2$ and $WS_{2(1-x)}Te_{2x}$, and the isostructural alloy $MoS_{2(1-x)}Te_{2x}$, for contrast. The phase diagram of $Mo_{1-x}W_xTe_2$ is found to closely match the one derived using time-consuming cluster expansion models (Duerloo and Reed, 2016). Similarly, the phase diagram of the isostructural alloy $MoS_{2(1-x)}Te_{2x}$ shows good agreement with that generated by Kang et al using cluster expansion models (Kang *et al.*, 2013a; Kang *et al.*, 2013b). The phase diagram of $Mo_{1-x}W_xTe_2$ has a large region of metastability with no unstable region and a cross-over from *2H* to *1T′* ground state at *x* ~ 0.33, making it a good candidate for phase engineering. In contrast, $WS_{2(1-x)}Te_{2x}$ has a large region where the alloys are unstable and are expected to segregate, and a cross-over from *2H* to *1T′* phase at $x = 0.9$. This indicates that the *1T′* phase is difficult to stabilize in $WS_2$, but a large concentration of Te can be added to $WS_2$, while maintaining the semiconducting *2H* phase. This method can be rapidly applied to the large space of TMDC alloys with various polytypes, and other related 2D materials such as MXenes.

**Results**

A recent paper by Hemmat et al. used the common-tangent construction method to generate equilibrium phase diagrams for 25 TMDC isostructural alloys with the *2H* phase (Hemmat *et al.*, 2020). In that work, chalcogens were restricted to S and Se because many tellurides have the *1T′* phase as their ground state. Here, we expand the method to generate phase diagrams for heterostructural TMDC alloys. We started with an isostructural TMDC alloy having Te, $MoS_{2(1-x)}Te_{2x}$. While bulk $MoTe_2$ can exist in either a distorted octahedral *1T′* phase or the triangular prismatic *2H* phase, monolayer $MoTe_2$ prefers the *2H* phase (Duerloo et al., 2014; Revolinsky and Beerntsen, 1966); and so does 2D $MoSe_2$. An equilibrium phase diagram of $MoS_{2(1-x)}Te_{2x}$ generated using the cluster expansion method shows that it exists in the *2H* phase for all compositions, *x* (Kang *et al.*, 2013a). We use this alloy as a benchmark to compare the phase diagram generated using a solution model fitted to the energy of disordered SQSs. The mixing enthalpy of $MoS_{2(1-x)}Te_{2x}$, calculated using SQS, is shown in Figure 1A. Because the mixing enthalpy is positive, the alloy is not miscible; that is, entropy is required to obtain a stable alloy in a single-phase.

While the enthalpy of some alloys and solutions can be modeled as a regular solution with a quadratic fit, the asymmetry of the enthalpy in Figure 1A clearly precludes this option. The regular solution model can be represented mathematically by the following equation:

$$\Delta H_{\text{mix}} = \Omega x(1-x). \tag{1}$$

Assuming nearest neighbor interactions, the fitting parameter, $\Omega$, represents the relative bonding strength between generic alloyed elements $A$ and $B$ in $A_{1-x}B_x$ compared to the $A$–$A$ and $B$–$B$ bonds. This can be expressed as:

$$\Omega = n\left(U_{AB} - \frac{1}{2}U_{AA} - \frac{1}{2}U_{BB}\right), \tag{2}$$

where $n$ is the coordination number and the $U$'s are bonding energy contributions to the enthalpy. See STAR Methods section for the derivation of Equation (2). The regular solution model for this system is illustrated in Figure 1A with a dashed line, which shows a poor fit. To capture the asymmetry of the enthalpy with respect to composition, a cubic sub-regular solution model is used. Such a model has the following general form:

$$\Delta H_{\text{mix}} = [\Omega_1(1-x) + \Omega_2 x]x(1-x). \tag{3}$$

In this form, the cubic sub-regular solution model can be seen as an average of two regular solution models weighted by $x$. A skewing of the enthalpy maximum to the right, i.e., $x > 0.5$, corresponds to $\Omega_2 > \Omega_1$. Similar to Equation (2), the fitting parameters in (3), $\Omega_1$ and $\Omega_2$, have a microscopic interpretation related to the relative cluster energies. For a triangular lattice, the fitting parameters are given by:

$$\Omega_1 = 6\left(U_{AB} - \frac{1}{2}U_{AA} - \frac{1}{2}U_{BB} + U_{AAB} - \frac{2}{3}U_{AAA} - \frac{1}{3}U_{BBB}\right) \tag{4}$$

and

$$\Omega_2 = 6\left(U_{AB} - \frac{1}{2}U_{AA} - \frac{1}{2}U_{BB} + U_{ABB} - \frac{1}{3}U_{AAA} - \frac{2}{3}U_{BBB}\right). \tag{5}$$

See STAR Methods section for the derivation of Equations (4) and (5).

To show that fitting a sub-regular solution model to SQSs also works for heterostructural alloys, we applied it to the TM-site and chalcogen-site alloys $Mo_{1-x}W_xTe_2$ and $WS_{2(1-x)}Te_{2x}$, respectively. The steps for creating equilibrium phase diagrams for heterostructural alloys follow a very similar recipe of common tangent construction with the additional complication that free energy curves must be determined for both endpoint phases. Furthermore, the appearance of the plotted mixing enthalpy takes on a qualitatively different form because it is calculated with respect to the ground state phase of the end points. Figure 2A shows the calculated mixing enthalpy of $Mo_{1-x}W_xTe_2$ in the *2H* and *1T'* phases with cubic fits. The mixing enthalpies are positive, indicating that the alloy is immiscible, but the curvature is positive everywhere, indicating that a single-phase alloy is at worst metastable, i.e., it does not have any unstable region. In contrast, the mixing enthalpy of $WS_{2(1-x)}Te_{2x}$, shown in Figure 2B, is seen to be comparatively high and has a negative curvature (concave down). This means that stabilization of a single phase will require a higher temperature and that there will be an unstable region where the alloy will be driven to phase segregation.

With analytical expressions for the enthalpy of mixing for the isostructural and heterostructural alloys, we have half of the expression for the Gibbs free energy. The other half requires an analytical expression for the entropy of mixing. The entropy of an alloy is a sum of configurational entropy and other sources of entropy such as from vibrations. If we assume that these other contributions extrapolate linearly from the end-member materials, the configurational entropy is the primary contribution to $\Delta S_{\text{mix}}$.(Manzoor et al., 2018) Because configurational entropy has a simple functional form (Equation (6)), we have an expression for the free energy. Figure 3A and Figure 3B show the free energies of isostructural $MoS_{2(1-x)}Te_{2x}$ and heterostructural $Mo_{1-x}W_xTe_2$, respectively, at 400 K. Through the common tangent construction, the dashed lines show the free energy of a phase segregated alloy. In the case of $MoS_{2(1-x)}Te_{2x}$, this indicates segregation into an S-rich *2H* phase and a Te-rich *2H* phase. For $Mo_{1-x}W_xTe_2$, the dashed line indicates segregation into an Mo-rich *2H* phase and a W-rich *1T'* phase. In alloys where the curvature of the free energy is negative, this region is further divided into a metastable and an unstable region corresponding to positive and negative curvature of the free energy, respectively. This boundary is demarked in Figure 3A with vertical dashes at the inflection points. Because the curvature of the mixing enthalpy of $Mo_{1-x}W_xTe_2$ is negative (Figure 2B), there is no inflection point in the free energy and no unstable region. Repeating this analysis of the free energy over a grid of temperature values gives boundaries dividing the different regions of stability which define the equilibrium phase diagram. The equilibrium phase diagrams for $MoS_{2(1-x)}Te_{2x}$, $Mo_{1-x}W_xTe_2$, and $WS_{2(1-x)}Te_{2x}$ are shown in Figure 4.

**Discussion**

We begin our discussion by addressing the nature of the cubic subregular solution model expressed in Equations (3) through (5). Just as the terms of the form $U_{XY}$ correspond to the bonding energy between species $X$ and $Y$, terms of the forms $U_{XYZ}$ correspond to a three-body contribution to the enthalpy that is supplementary to the three corresponding two-body terms. Equations (4) and (5) can be seen as corrections to the coefficient defined by Equation (2). The additional correction terms themselves take on a similar form to Equation (2), corresponding to a difference between an alloy energy term and a weighted average of the pure elemental terms. Therefore, a skewing of the enthalpy maximum to the right will correspond to a greater relative energy contribution from *A-B-B* triplets compared to *A-A-B* triplets. An opposite skew will correspond to an opposite relationship between the triplet energies.

We postulate that the requirement for a cubic fit of the enthalpy of $MoS_{2(1-x)}Te_{2x}$ is a natural consequence of the coordination of the *2H* phase. While the regular solution model is derived from 2-body energy contributions, cubic terms become important when 3-body interactions cannot be ignored. Figure 1B demonstrates why 3-body interactions are naturally important in quasibinary alloying of the *2H* and *1T'* phases compared to strictly binary alloys. The solid arrows represent an interaction between two alloyed sites. In quasibinary TMDC alloys, either the TM site or the chalcogen site can be alloyed. In either case, the interaction is facilitated through an intermediate chalcogen pair or transition metal, respectively. Connected to this intermediate is another TM or

chalcogen — that we refer to as the tertiary site. The occupation of this tertiary site will affect the bonding properties of the intermediate site, making 2-body interactions insufficient as visualized by the dashed arrows. This argument is not restricted to TMDCs: a large class of ceramics have cation-cation interactions mediated through an anion coordinated with three of more cations or vice versa. Some examples include perovskites, wurtzite-structure materials, and MXenes.

In the case of *1T'* phase, there are many types of nearest neighbor 3-body clusters, but this does not change the argument for including a cubic term in the enthalpy. In the case of $MoS_{2(1-x)}Te_{2x}$, Figure 1A shows the mixing enthalpy peak is shifted to the right, indicating that $\Omega_2 > \Omega_1$. From Equations (4) and (5), this asymmetry is determined to be caused by the relatively high energy of S-Te-Te clusters compared to S-S-Te clusters.

We now focus on the equilibrium phase diagrams shown in Figure 4. The equilibrium phase diagram for $MoS_{2(1-x)}Te_{2x}$ is shown in Figure 4A. Because the alloy is isostructural, the phase diagram is relatively straightforward with a single stable region and a single unstable region with two metastable regions connected at the critical point. The temperature at the critical point is known as the miscibility temperature, the temperature above which the alloy is stable at all compositions. We note that the slight asymmetry of the enthalpy in Figure 1A led to more substantial asymmetry in the phase diagram, making Te-rich alloys more difficult to synthesize in a single phase. The predicted asymmetries in the phase diagram of some *2H* TMDC alloys was experimentally verified in a previous work (Hemmat *et al.*, 2020). This phase diagram generated with the SQS method is nearly identical in miscibility temperature and asymmetry to previous cluster expansion-based works where the metastable region was omitted (Kang *et al.*, 2013a; Kang *et al.*, 2013b).

The equilibrium phase diagram of $Mo_{1-x}W_xTe_2$ shown in Figure 4B is more complicated because of its heterostructural nature. One simplification is that there is no unstable region due to the positive curvature of the mixing enthalpy (Figure 2B) for all *x*. The large metastable region and the near-equimolar cross-over concentration makes this a promising material for phase engineering. These results are qualitatively similar to a work by Duerloo and Reed using a cluster expansion method that shows a similar cross-over concentration and a similar range of compositions over which the *2H* and *1T'* phases are stable (Duerloo and Reed, 2016). While Figure 2B shows a temperature-independent cross-over concentration, Duerloo and Reed's work has a cross-over region that converges to *x*=0, indicating the stabilization of pure $MoTe_2$ in the 1T phase. This discrepancy is because they included vibrational contributions to the free energy. This omission increases the uncertainty of pur equilibrium phase diagrams as the temperature approaches the phase transition temperatures of the pure TMDC components. See Figure S1 in the Supplemental Information for a comparison.

Lastly, the phase diagram for $WS_{2(1-x)}Te_{2x}$ is shown in Figure 4C. This heterostructural alloy produces the most complex phase diagram of the three because of its unstable region. Because the curvature of the mixing enthalpies corresponding to the two phases are different, the unstable region is defined piecewise with an ambiguous crossover region. $WS_2$ strongly favors the *2H* phase with the DFT-calculated difference in energy of *2H* and *1T'* phase, $\Delta E_{(1T'-2H)} = 542$ meV/f.u. Therefore, the crossover region for the alloy is at a very high Te content, indicating that the *1T'* phase of the alloy is difficult to realize experimentally. This could possibly be overcome by using Se instead of S because $WSe_2$ is less energetically opposed to the *1T'* phase with $\Delta E_{(1T'-2H)} = 279$ meV/f.u.. While to our knowledge this is the first generation of a heterostructural phase diagram for $WS_{2(1-x)}Te_{2x}$, a phase diagram for this alloy assuming only the *2H* phase shows agreement with our unstable region (Kang *et al.*, 2013a; Kang *et al.*, 2013b). Recent work has shown that the generation of equilibrium phase diagrams can successfully lead to the discovery of new miscible 2D alloys and in guiding the synthesis of immiscible 2D alloys (Hemmat *et al.*, 2020). Furthermore, heterostructural 2D alloys have become a promising platform for tuning properties and phase engineering. We have presented a method for generating these phase diagrams that is standardized and only requires a few calculations of moderately sized SQS's. We showed that this method is applicable to both isostructural and heterostructural alloys and it produces results that agree with previous works using more computationally intensive methods. This method of generating equilibrium phase diagrams has the potential to accelerate the expanding field of 2D alloys by providing guidance for the synthesis of 2D materials, such as high-entropy alloys of TMDCs and MXenes (Cavin *et al.*, 2021; Nemani et al., 2021).

**Limitations of Study**

The accuracy of this work is limited by multiple factors. The accuracy of the predictions is affected by the approximate nature of the exchange-correlation functionals used in DFT. We have also neglected the vibrational and electronic contributions to entropy . The inclusion of these would likely increase entropy, lowering the temperature of various phase boundaries. Our work therefore describes upper bounds for the phase transition barriers. Vibrational contributions to the enthalpy have also been neglected. These can have significant impact on free energy differences between competing phases at high temperatures (Duerloo and Reed, 2016).

**Acknowledgements**

This work was supported by the National Science Foundation (NSF) through a DMREF grant CBET-1729787. This work used computational resources of the Extreme Science and Engineering

Discovery Environment (XSEDE), which is supported by NSF ACI-1548562. The authors thank Dr. Sung Beom Cho for useful discussions.

**Author Contributions**

J.C. and R.M. conceptualized the project. J.C. developed the software and methodology used to carry out the project under the supervision of R.M. J.C. created the data visualizations. J.C. wrote the first draft, and reviewing and editing was done by both J.C. and R.M.

**Declaration of Interests**

John Cavin is currently a postdoctoral scholar at Northwestern University. The authors declare no competing interests.

**Main Figure Title and Legends**

Figure 1. Enthalpy of mixing of $MoS_{2(1-x)}Te_{2x}$ fitted to different solution models and atomic models justifying the need for a sub-regular solution model. (A) Mixing enthalpy from DFT calculation of SQS's at various compositions, shown as datapoints, with a quadratic regular solution model fit (dashed line) and cubic sub-regular solution model fit (solid line). (B) Visualizations of the *2H* and *1T'* phases showing how 3-body clusters are essential in 2D TMDCs. Transition metals are shown in purple and chalcogen atoms are shown in gold. Solid arrows indicate direct interactions between two sites, while dashed arrows indicate indirect interactions through a third site.

Figure 2. Enthalpy of mixing for two heterostructural TMDC alloys with cubic sub-regular solution model fits. (A) $Mo_{1-x}W_xTe_2$, a TM-site heterostructural alloy. (B) $WS_{2(1-x)}Te_{2x}$, a chalcogen-site heterostructural alloy. The vertical dashed line denotes the composition where the stable phase changes.

Figure 3. Free energy of an isostructural and a heterostructural alloy at 400 K. (A) $MoS_{2(1-x)}Te_{2x}$, an isostructural alloy. Vertical black dashes indicate the boundary between the unstable and metastable regions (inflection points). (B) $Mo_{1-x}W_xTe_2$, a heterostructural alloy. The colored dash lines denote the free energy of a mixed phase.

Figure 4. Equilibrium phase diagrams for three representative TMDC alloys. (A) $MoS_{2(1-x)}Te_{2x}$, an isostructural alloy. (B) $Mo_{1-x}W_xTe_2$, a heterostructural alloy without an unstable region. (C) $WS_{2(1-x)}Te_{2x}$, a heterostructural alloy with an unstable region. Vertical dashed lines in (B) and (C) correspond to the composition where the stable phase changes. See also Figure S1.

**STAR Methods**

**RESOURCE AVAILABILITY**

*Lead contact*

Further information and requests for resources should be directed to and will be fulfilled by the lead contact, Rohan Mishra (rmishra@wustl.edu).

*Materials availability*

This study did not generate new unique reagents.

*Data and code availability*

- Structure files for the alloy SQS's are provided. The DOI is listed in the key resources table.
- Original code has been deposited on GitHub through Zenodo. The DOI is listed in the key resources table.
- Any additional information required to reanalyze the data reported in this paper is available from the lead contact upon request.

**METHOD DETAILS**

*Generating Equilibrium Phase Diagrams*

A standard method for generating equilibrium phase diagrams is by using semi-grand canonical ensemble lattice Monte Carlo simulations to determine phase boundaries directly through thermodynamic integration (Walle and Asta, 2002). These Monte Carlo simulations are typically carried out through the cluster expansion method (Sluiter *et al.*, 1990; van de Walle *et al.*, 2002). Cluster expansions decompose the internal energy of an alloy into a sum over cluster contributions. This method is useful if the sum converges quickly with respect to the maximum cluster size included. By performing *ab initio* DFT calculations on many relatively small supercells with varying alloy configurations, the energy of 1-, 2-, and 3-body clusters can be fit with least squares regression. These simulations can individually be quite expensive and cumulatively are even more so as many are required to perform the thermodynamic integration.

A more straightforward method to generate phase diagrams is to use the common tangent construction and metastability analysis on analytical expressions of the Gibbs free energies of all the relevant alloy phases. Specifically, the free energy of mixing of the alloy with respect to its end members, $\Delta G_{mix}$, is required for all intermediate compositions, $x \in (0,1)$. The Gibbs free energy is composed of two terms:

$$\Delta G_{mix} = \Delta H_{mix} - T\Delta S_{mix}, \qquad (6)$$

where $\Delta H_{mix}$ and $\Delta S_{mix}$ are the enthalpy and the entropy of mixing, respectively, and $T$ is the temperature. The mixing entropy, $\Delta S_{mix}$, can be approximated with the configurational entropy:

$$\Delta S = -k_B[x \ln x + (1-x)\ln(1-x)]. \qquad (7)$$

Here, $k_B$ is the Boltzmann constant. One method to get an analytical form for $\Delta H_{mix}$ is to use a cluster expansion as outlined before. Replacing the correlation functions present in a cluster expansion sum with the statistical values in a random alloy gives a polynomial expression for $\Delta H_{mix}$. The degree of the polynomial directly corresponds to the largest cluster size included in the expansion where using up to $N$-element clusters corresponds to a $N^{th}$ degree polynomial. This procedure has been used before for constructing phase diagrams of binary TMDC alloys (Duerloo and Reed, 2016; Kang *et al.*, 2013a).

An alternative method to get an expression for mixing enthalpy is to directly perform a polynomial fit — that emulates a thermodynamic solution model — to calculated enthalpies at various compositions. A brute force approach is to create large supercells with pseudorandom occupation of sites corresponding to many composition values and fitting the results to a polynomial as has been applied for bulk semiconductor alloys (Holder *et al.*, 2017). A more convenient method is to use special quasirandom structures (SQS's) (Wei *et al.*, 1990a; Zunger *et al.*, 1990), which are relatively small supercells having site occupations that closely mimic the distribution of clusters found in a random alloy with the same composition. More precisely, they replicate the spin-variable correlation functions that serve as the basis for cluster expansions. Therefore, properties such as enthalpy expressions that are well represented by truncated cluster expansions can be calculated from SQS's on the condition that the SQS's are large enough to accurately represent the clusters corresponding to the truncated expansion. This was demonstrated by Zunger et al. for mixing enthalpy and band gaps in their original paper (Zunger et al., 1990). Because enthalpy expressions can be determined with a relatively small number of moderately-sized SQS's, we chose this method. The generation of SQS's was facilitated by a reverse monte Carlo method implemented in the Alloy Theoretic Automated Toolkit (ATAT) (van de Walle *et al.*, 2002). Because SQS's suitably mimic the clusters of a theoretical alloy, only one calculation is required per composition value, and because they are significantly smaller than pseudorandomly occupied supercells required for similar accuracy, they are advantageous in terms of computation time (Wei *et al.*, 1990a; Zunger *et al.*, 1990). These considerations make SQS's a powerful tool for scalable generation of equilibrium phase diagrams.

With an expression for the free energy, the composition space can be divided into different regions of stability for a fixed temperature. This procedure is repeated over a grid of temperatures to fully develop the phase boundaries and determine the full equilibrium phase diagram. For an isostructural alloy, there can be three regions: stable, metastable, and unstable. The stable region is the set of compositions where the single-phase alloy is on the convex hull. Outside of this region, metastability is determined by the curvature of the free energy, i.e., $\frac{d^2}{dx^2}\Delta G_{\text{mix}}$. When the curvature is positive and the free energy itself is not on the convex hull, the free energy of the alloy is lower than the average energy of two infinitesimally close compositions on the adjacent side of the alloy. Therefore, in this region, the single-phase alloy is metastable in the sense that it is stable to small fluctuations. However, if the curvature of the free energy is negative, infinitesimal phase decompositions *lower* the free energy compared to the single-phase alloy. Therefore, the metastable and unstable regions correspond to positive and negative curvatures of the free energy, respectively, outside the convex hull. The binodal boundary between these regions is demarked by an inflection point. This treatment is easily generalized to multiple structural phases by including multiple free energy surfaces in the convex hull analysis leading to different stable regions for each morphology.

*Regular Solution Model*

Consider a crystalline material consisting of a single element $A$. Using the cluster expansion method, the enthalpy of such a material can be expressed as a sum over effective cluster interactions (van de Walle *et al.*, 2002). For simplicity, we will only consider 1-body and nearest neighbor 2-body interactions. Therefore, the enthalpy of material $A$ is given by the following equation:

$$H_A^{\text{tot}} = N_A U_A + N_{AA} U_{AA}. \tag{8}$$

Here, $N_A$ and $N_{AA}$ are the number of $A$ atoms and $A$-$A$ bonds, respectively. $U_A$ and $U_{AA}$ are then the energy contributions to the enthalpy corresponding to effective cluster interactions. This expression can be simplified further by introducing the coordination number, $n$:

$$H_A^{\text{tot}} = N_A \left( U_A + \frac{n}{2} U_{AA} \right). \tag{9}$$

Now, consider a random solid solution of elements $A$ and $B$. The total enthalpy of such a system is given by the following equation:

$$H_{AB}^{\text{tot}} = N_A U_A + N_B U_B + N_{AA} U_{AA} + N_{BB} U_{BB} + N_{AB} U_{AB}. \tag{10}$$

Here, most of the terms follow directly from Equation (8) with the addition of $N_{AB}$ and $U_{AB}$, which correspond to the quantity and energy of the $A$-$B$ bonds, respectively. Let $x$ be a real number between 0 and 1 that denotes the concentration of element $B$:

$$x = \frac{N_B}{N_A + N_B} = \frac{N_B}{N}. \tag{11}$$

Here, we defined $N$ as the total number of atoms, i.e. the sum of the quantities of $A$ and $B$ atoms. The quantity that we wish to determine is the mixing enthalpy of the alloy,

$$\Delta H_{\text{mix}}^{\text{tot}} = H_{AB}^{\text{tot}} - (1-x)H_A^{\text{tot}} - xH_B^{\text{tot}}. \tag{12}$$

More specifically, we are interested in the mixing enthalpy per atom. Dividing Equation (12) by $N$ and expressing $x$ dependence explicitly gives

$$\Delta H_{\text{mix}}(x) = H_{AB}(x) - (1-x)H_A - xH_B. \tag{13}$$

It is important to note that in the direct substitution of Equations (8) and (10) into Equation (13), the $N_A$ and $N_{AA}$ in equation correspond to a pure material consisting of $N$ atoms of species $A$, not $(1-x)N$ atoms. All that remains in evaluating Equation (13) is to find expressions for variables of the form $N_X$ and $N_{XY}$, where $N_X$ is the number of atoms of type $X$, and $N_{XY}$ is the number of bonds between X–Y. This can be done using conditional probability. For a single site, this is trivial:

$$P(A|x) = 1 - x \tag{14}$$

and

$$P(B|x) = x. \tag{15}$$

For a pair of neighboring sites, the probabilities are given by terms in a binomial expansion:

$$P(AA|x) = (1-x)^2, \tag{16}$$

$$P(AB|x) = 2x(1-x), \tag{17}$$

and

$$P(BB|x) = x^2. \tag{18}$$

From these expressions, the $N_X$ and $N_{XY}$ terms can be multiplying the probabilities above by the total numbers of atoms, $N$, or the total number of bonds, $\frac{n}{2}N$. Performing these substitutions for Equation (10) gives the following:

$$H_{AB}^{tot} = N\left\{(1-x)U_A + xU_B + \frac{n}{2}[(1-x)^2 U_{AA} + x^2 U_{BB} + 2x(1-x)U_{AB}]\right\}. \tag{19}$$

Substituting Equations (19) and (9) into equation (12) gives

$$\Delta H_{mix}^{tot} = \frac{Nn}{2}\{[(1-x)^2 - (1-x)]U_{AA} + (x^2 - x)U_{BB} + 2x(1-x)U_{AB}\}. \tag{20}$$

Simplifying and dividing by $N$ gives

$$\Delta H_{mix} = n\left(U_{AB} - \frac{1}{2}U_{AA} - \frac{1}{2}U_{BB}\right)x(1-x), \tag{21}$$

the same result as Equations (1) and (2). This derivation shows that a simple, one-parameter quadratic expression for the mixing enthalpy of a solid solution follows from a model that assumes that the internal energy can be written as a sum of one- and two-body interactions. This model is referred to as the regular solution model.

*Sub-regular Solution Model*

The regular solution model is derived from a truncated cluster expansion model. The regular solution model keeps only one- and two-body interactions, but hypothetically, higher order interactions can be kept. Such models are called sub-regular solution models. Here, we derive the expression for the cubic sub-regular solution model, but the derivation can be generalized to higher orders. Just as the regular solution model was derived using one- and two-body interactions, we will derive the cubic sub-regular solution model by also including three-body interactions.

Analogous to Equations (8) and (10), the enthalpies of a pure material and a solid solution can be expressed, respectively, through

$$H_A^{\text{tot}} = H_A^{\text{reg}} + N_{AAA} U_{AAA} \qquad (22)$$

and

$$H_{AB}^{\text{tot}} = H_{AB}^{\text{reg}} + N_{AAA} U_{AAA} + N_{AAB} U_{AAB} + N_{ABB} U_{ABB} + N_{BBB} U_{BBB}. \qquad (23)$$

Variables of the form $N_{XYZ}$ and $U_{XYZ}$ naturally correspond to quantities and energies of 3-body clusters. The terms $H_A^{\text{reg}}$ and $H_{AB}^{\text{reg}}$ are defined by Equations (8) and (10) for brevity. Because 3-body clusters have more degrees of freedom in their specific shape, we will heretofore assume a 2D triangular lattice rather than a general lattice with arbitrary coordination. This is because such a lattice is relevant to the TMDC alloys we are studying, and because the functional form of the result is independent of the choice of the lattice. Furthermore, we will only consider the 3-body energy contributions of equilateral triangles of nearest neighbors. With this in mind, Equation (22) can be expanded as

$$H_A^{\text{tot}} = N(U_A + 3U_{AA} + 2U_{AAA}). \qquad (24)$$

Figure S2 shows an excerpt of a triangular lattice with colored shapes representing different $n$-body energy contributions per atom. The number of red dots, green lines, and blue triangles correspond to the coefficients in Equation (24).

Just as the $N_X$ and $N_{XY}$ variables can be determined through conditional probabilities, so too can the $N_{XYZ}$ variables. All four such variables can by expressed in the following condensed formula:

$$P(ijk|x) = \binom{3}{M_A}(1-x)^{M_A} \times x^{(3-M_A)}. \tag{25}$$

Here, $M_A$ is the number of variables out of $X$, $Y$, and $Z$ that equal $A$. Plugging Equations (22) through (25) into Equation (12) gives

$$\Delta H_{AB}^{tot} = \Delta H_{mix}^{reg} \tag{26}$$
$$+ 2N\{[(1-x)^3 - (1-x)]U_{AAA} + (x^3 - x)U_{BBB}$$
$$+ 3x(1-x)^2 U_{AAB} + 3x^2(1-x)U_{ABB}\}.$$

This expression can be simplified by factoring out $x(1-x)$ out of the parenthetical term:

$$\Delta H_{AB}^{tot} = \Delta H_{mix}^{reg} \tag{27}$$
$$+ 2N[(x-2)U_{AAA} - (x+1)U_{BBB} + 3(1-x)U_{AAB}$$
$$+ 3xU_{ABB}]x(1-x).$$

It is particularly instructive to group the second term as a sum of $(1-x)$ and $x$ terms:

$$\Delta H_{AB}^{tot} = \Delta H_{mix}^{reg} \tag{28}$$
$$+ 2N[(3U_{AAB} - 2U_{AAA} - U_{BBB})(1-x)$$
$$+ (3U_{ABB} - U_{AAA} - U_{BBB})x]x(1-x).$$

Dividing by $N$ and substituting Equation (21) gives the familiar expression for the sub-regular solution model expressed in Equation (3):

$$\Delta H_{mix} = [\Omega_1(1-x) + \Omega_2 x]x(1-x), \tag{29}$$

with $\Omega_1$ and $\Omega_2$ given by Equations (4) and (5).

*Computational Details*

Disordered 2D alloys were studied using $6 \times 6$ and $3 \times 6$ supercells of the primitive cell of $2H$-phase and $1T'$-phase TMDCs, respectively. Structure files can be found at a link provided in the Key Resources Table (Cavin, 2022). Slab models with vacuum along the out-of-plane direction were used to simulate 2D layers. To eliminate interaction between image 2D layers, a vacuum spacing of >15 Å in the out-of-plane direction was used. Total energies were calculated using DFT as implemented in the Vienna Ab-initio Simulation Package (VASP) using the Perdew-Burke-Ernzerhof exchange-correlation functional (Kresse and Furthmüller, 1996; Perdew *et al.*, 1996). For the SQS's, geometric relaxation was conducted at only the Γ-point in reciprocal space allowing for in-plane relaxation of lattice parameters. A subsequent static calculation for the electronic structure was performed using a Γ-centered $3 \times 3 \times 1$ $k$-points mesh generated using the Monkhorst-Pack method (Monkhorst and Pack, 1976). Pure TMDCs were studied using a $k$-points grid of $8 \times 8 \times 1$ for geometry optimization and $24 \times 24 \times 1$ for static calculations. A kinetic energy cutoff of 450 eV was used for all the calculations.